\documentclass[preprint, colorlinks, linkcolor=refcol, citecolor=refcol, urlcolor=refcol, a4paper, fleqn]{cas-dc}
\usepackage[english]{babel}
\usepackage[utf8]{inputenc}
\usepackage{xcolor}
\definecolor{refcol}{RGB}{178,34,34}%{178,34,34}%{0,0,205}
\usepackage[numbers,sort&compress]{natbib}
\graphicspath{{./figures/}}
\newcommand{\be}{\begin{equation}}
\newcommand{\ee}{\end{equation}}

\newcommand{\tr}{\text{tr}}

\newcommand{\eff}{\mathrm{eff}}
\newcommand{\gen}{\mathrm{gen}}
\newcommand{\phys}{\mathrm{phys}}
\definecolor{red}{rgb}{1,0,0}
\newcommand{\red}[1]{{\color{red} #1}} 
\begin{document}
\shorttitle{Chiral anomaly: from vacuum to Columbia plot}
\shortauthors{F.~Giacosa~at~al}

\title [mode = title]{Chiral anomaly: from vacuum to Columbia plot}
\tnotemark[T1]
\tnotetext[T1]{Contribution to XQCD 2025 Wroclaw}

\author[L1,L2]{\color{black} Francesco Giacosa}[orcid=0000-0003-3735-7620]
\ead{francesco.giacosa@ujk.edu.pl}

\author[L3,L4]{\color{black} {Gy}őző Kovács}[orcid=0000-0003-3735-7620]
%\ead{gyozo.kovacs@uwr.edu.pl}

\author[L4]{\color{black} Péter Kovács}[orcid=0000-0003-3735-7620]
%\ead{kovacs.peter@hun-ren.wigner.hu}
   
\author[L5]{\color{black} Robert D. Pisarski}[orcid=0000-0003-3735-7620]
%\ead{pisarski@bnl.gov}

\author[L6,L7]{\color{black} Fabian Rennecke}[orcid=0000-0003-3735-7620]
%\ead{fabian.rennecke@theo.physik.uni-giessen.de}

\affiliation[L1]{organization={Institute of Physics, Jan Kochanowski University},
            addressline={ulica Uniwersytecka 7},
            city={Kielce},
            postcode={ P-25-406},
            country={Poland}}
\affiliation[L2]{organization={Institute for Theoretical Physics, Goethe-University},
            addressline={Max-von-Laue-Straße 1},
            city={Frankfurt am Main},
            postcode={D-60438},
            country={Germany}}
\affiliation[L3]{organization={University of Wrocław},
            addressline={Plac Maxa Borna},
            city={Wrocław},
            postcode={50204},
            country={Poland}}
\affiliation[L4]{organization={HUN-REN Wigner Research Centre for Physics},
            addressline={Konkoly-Thege Mikos ut 29-33},
            city={Budapest},
            postcode={1121},
            country={Hungary}}
\affiliation[L5]{organization={Department of Physics, Brookhaven National Laboratory},
            addressline={Plac Maxa Borna},
            city={Upton, NY},
            postcode={11973},
            country={USA}}
\affiliation[L6]{organization={Institute for Theoretical Physics, Justus Liebig University Giessen},
            addressline={Heinrich-Buff-Ring 16},
            city={Giessen},
            postcode={35392},
            country={Germany}}
\affiliation[L7]{organization={Helmholtz Research Academy Hesse for FAIR (HFHF)},
            addressline={Campus Giessen},
            city={Giessen},
            postcode={35392},
            country={Germany}}

%% Keywords
%\begin{keywords}
%Finite size effects \sep QCD phase diagram \sep Baryon fluctuations \sep Critical endpoint \sep Quark-meson model
%\end{keywords}

\date{\today} % Leave empty to omit a date

\begin{abstract}
We use a low-energy effective approach, the extended linear sigma model, to study realizations of the $U(1)_A$ anomaly with different operators, linear and quadratic in the 't Hooft determinant.
After discussing the parameterization in agreement with vacuum's phenomenology, we investigate the influence of these different anomaly terms on the Columbia plot: the square of the 't Hooft determinant favors a cross-over for small quark masses. Finally, we also discuss the extension of the 't Hooft determinant to cases in which different mesonic multiplets interact with each other. Novel chiral anomalous interaction terms involving excited (pseudo)scalar states, pseudovector, and pseudotensor mesons are expressed via a mathematical extension of the determinant, denoted as a \textit{polydeterminant}.   
\end{abstract}

%% Keywords

\maketitle

\hypersetup{
pdftitle={Chiral anomaly: from vacuum to Columbia plot},
pdfsubject={XQCD proceedings},
pdfauthor={F. Giacosa, G. Kovacs, P. Kovacs, R. Pisarski, F. Rennecke}
}

%%%%%%%%%%%%%%%%%%%%%%%%%%%%%%%%%%%%%%%%%%
%%%%%%%%%%%%%%%%%%%%%%%%%%%%%%%%%%%%%%%%%%
\section{Introduction} \label{sec:intro}
%%%%%%%%%%%%%%%%%%%%%%%%%%%%%%%%%%%%%%%%%%
%%%%%%%%%%%%%%%%%%%%%%%%%%%%%%%%%%%%%%%%%%

The chiral (or axial) anomaly is an important feature of Quantum Chromodynamics (QCD): this  $U(1)_A$ symmetry of the classical Lagrangian in the chiral limit (the limit in which the bare quark masses vanish) is broken by gluonic quantum fluctuations. 

It is well known that the chiral anomaly has an important impact on vacuum physics: the mass of the $\eta^{\prime}(958)$ meson receives a large contribution, making it not a quasi-Goldstone boson, not even in the chiral limit \cite{tHooft:1976rip,tHooft:1986ooh}. In turn, this anomaly is also responsible for the very good realization of isospin symmetry at the level of the pions \cite{Pisarski:1983ms,Gross:1979ur}. (For a recent unexpected breaking of isospin symmetry far beyond expectations, see Refs. \cite{NA61SHINE:2023azp,Brylinski:2023nrb}). 

The effective description of the chiral anomaly in the vacuum takes place via suitable determinant-like terms \cite{tHooft:1976rip,tHooft:1986ooh,Rosenzweig:1981cu,Witten:1980sp}. In fact, the determinant is tailor-made to preserve $SU(3)_L \times SU(3)_R$ chiral symmetry but to break the $ U(1)_A$ classical axial symmetry, reflecting the QCD chiral (or axial) anomaly. 
In Sec. 2.1 we recall how these features are implemented in the context of the extended linear sigma model with (axial-)vector mesons (eLSM, see e.g. Refs. \cite{Parganlija:2012fy,Giacosa:2024epf}).
In particular, we consider an additional term proportional to the square of the usual 't Hooft determinant. 
Next, in Sec. 2.2 we describe the vacuum parametrization of these anomaly terms in agreement with the constraints provided by vacuum physics (masses and decays listed in the PDG \cite{ParticleDataGroup:2024cfk}).
Besides these determinant terms, in Sec. 2.3 we discuss other anomalous interactions are possible by using a generalization of the determinant, named polydeterminant in Ref. \cite{Giacosa:2025dvu}. This object allows to introduce interaction terms for distinct chiral multiplets and for glueball states. 

The chiral anomaly is also important at nonzero
temperature $T$, since it may affect the order of the chiral phase transition. For this reason, it was extensively studied in the past, e.g. Refs. \cite{Pisarski:1983ms,Aoki:2006we,Butti:2003nu, Resch:2017vjs, Fejos:2022mso} and refs therein. 
Recently, in Ref. \cite{Pisarski:2024esv} under general terms and in Ref. \cite{Giacosa:2024orp} within the eLSM as a concrete approach, the role of the previously mentioned chiral anomalous terms was studied, in particular in connection with the Columbia plot. To this end, we recall that for three massless quark flavors (origin of the Columbia plot), a first-order transition has been predicted irrespective of the fate of $U(1)_A$ \cite{Resch:2017vjs}. In lattice QCD, a second-order transition for $N_f\leq 6$ massless quark flavors is obtained \cite{Cuteri:2021ikv}. For $N_f = 3$, other results determine stringent upper bounds for critical pion masses for second-order transition \cite{Bazavov:2017xul,Kuramashi:2020meg,Dini:2021hug}, see also Dyson-Schwinger second-order outcome of Ref. \cite{Bernhardt:2023hpr}. Here, we show how the novel anomalous term, the square of the 't Hooft determinant, strongly affects the Columbia plot, enlarging the cross-over region.  
The status is summarized in Sec. 3, where a different implementation of the Columbia plot than in Ref. \cite{Giacosa:2024orp} is shown.

Finally, in Sec. 4 we conclude and present some outlooks.

\section{Anomalous interactions and the parametrisation} \label{sec:LSM}
%%%%%%%%%%%%%%%%%%%%%%%%%%%%%%%%%%%%%%%%%%
%%%%%%%%%%%%%%%%%%%%%%%%%%%%%%%%%%%%%%%%%%
\subsection{'t Hooft anomaly terms within the eLSM} \label{subsec:thooft}
%%
%%%%%%%%%%%%%%%%%%%%
\begin{figure*}[t]
    \centering
    \includegraphics[width=.42\textwidth]{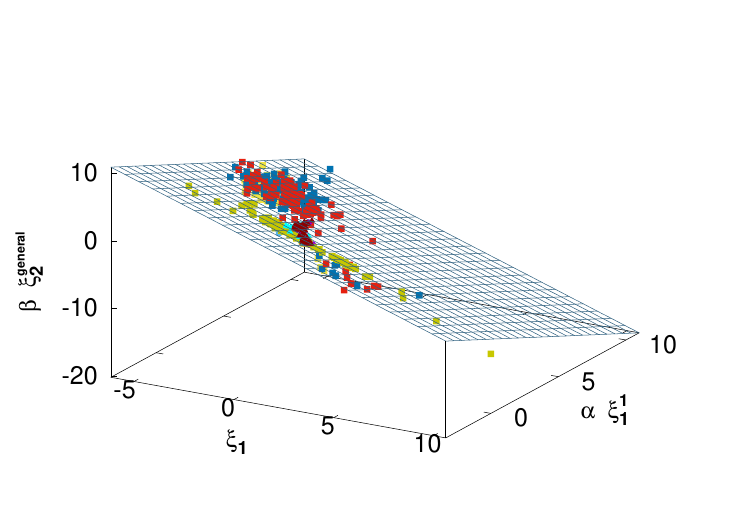}
    \includegraphics[width=.42\textwidth]{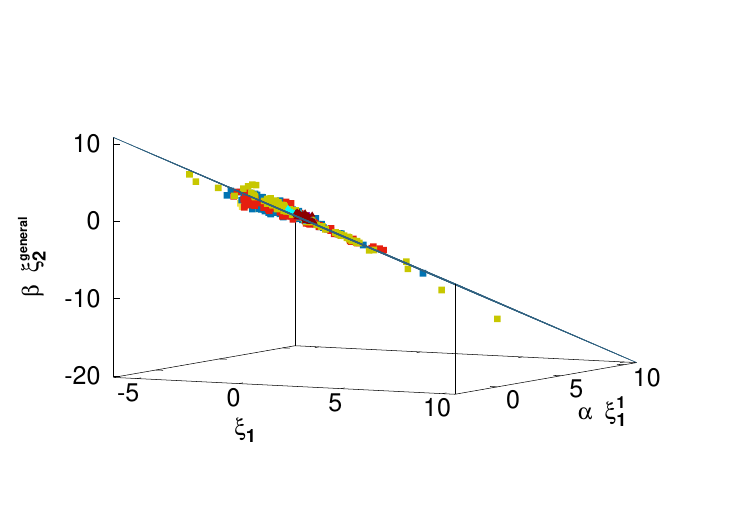}
    \includegraphics[width=.147\textwidth]{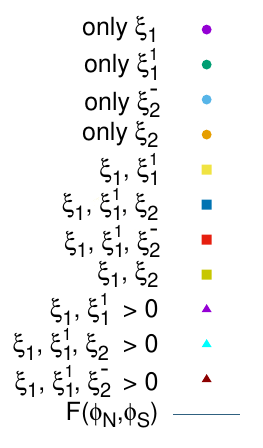} \hfill
    \caption{The different parameter sets in the parameters subspace spanned by the $\xi_1$, $\xi_1^1$, and $\xi_2^\mathrm{gen}=\xi_2,\xi_2^\pm$ are aligned on the $F(\phi_N,\phi_S)$ surface $\xi_1 + \alpha\,\xi_1^1+\beta\,\xi_2^\gen=\bar\xi_\eff $ determined by the constraint in Eq.~\eqref{Eq:chieff}. The numerical values are $\alpha=(0.13^2 + 0.14^2)/2$ and $\beta=0.13^2\cdot0.14/\sqrt2$, while $\bar\xi_\eff=1.4$ is used, but it is not fixed during the parametrization procedure.}
    \label{Fig:parametrizations}
\end{figure*}
%%%%%%%%%%%%%%%%%%%%
%%

In this work, we use the eLSM introduced for $N_f=2+1$ in Refs.~\cite{Parganlija:2012fy, Janowski:2014ppa} and applied at nonzero temperature and chemical potential in Refs.~\cite{Kovacs:2016juc, Kovacs:2021kas}. 
For a recent review, see \cite{Giacosa:2024epf}.
Continuing the work of Ref.~\cite{Giacosa:2024orp}, we discuss the effect of different $U(1)_A$ anomalous terms on the vacuum's parametrization and on the Columbia plot, while highlighting different aspects of our results. 

The Lagrangian of the eLSM is expressed in terms of a (pseudo)scalar field $\Phi$ (that contains pions, kaons, $\eta_{N,s}$ and their scalar chiral partners) as well as (axial-)vector fields. Under  $SU(3)_L \times SU(3)_R$ chiral transformations,  $\Phi \rightarrow U_L \Phi U_R^{\dagger}$, where $U_{L/R}$ are special unitary matrices. This transformation is called heterochiral \cite{Giacosa:2017pos}.

The Lagrangian includes a classical chirally invariant part $\mathcal{L}_{\rm cl}$ and additional terms that account for the explicit symmetry breaking $\mathcal{L}_{\rm ESB}$ and the axial anomaly $\mathcal{L}_{\rm qu}$. The term  $\mathcal{L}_{\rm ESB}$ is simply given by $\mathcal{L}_{\rm ESB}=\tr(H\Phi)$, which gives rise to $h_N\phi_N+h_S\phi_S$ when the fields are taken to their nonstrange and strange vacuum expectation values, $\phi_N$ and $\phi_S$ respectively.
For the term $\mathcal{L}_{\rm qu}$, multiple operators satisfy the expected symmetry properties. Generally, one can write
\begin{align} \label{Eq:Lqu}
    \begin{split}
        \mathcal{L}_{\rm qu} &= -\xi_1 \big(\det\Phi + \det\Phi^\dagger\big)\\
        &\quad - \xi_1^1 {\rm tr}\big( \Phi^\dagger\Phi\big) \big(\det\Phi + \det\Phi^\dagger\big)\\
        &\quad -\xi_2 \Big[\big(\det\Phi\big)^2 + \big(\det\Phi^\dagger\big)^2\Big]\,.
    \end{split}
\end{align}
The first term is the mesonic 't~Hooft determinant related to fluctuations of field configurations with topological charge $Q=1$ \cite{tHooft:1976snw}. The second term mixes anomalous and classical invariant contributions. The last, sixth-order term is generated by fluctuations with $Q=2$ \cite{Pisarski:2019upw}. 
Instead of this term, one can use either of
\be 
\mathcal{L}_{\rm qu}^{Q=2,\,\pm}= - \xi_2^\pm \Big[\big(\det\Phi\big) \pm \big(\det\Phi^\dagger\big)\Big]^2 \,,
\ee 
which differs only by a non-anomalous $\sim \det\Phi \det\Phi^\dagger$ invariant contribution. While the Lagrangian in Eq.~\eqref{Eq:Lqu} is the most general anomalous contribution with mass dimension $d_m\leq6$ operators, for completeness, we investigate the parametrization including $\xi_2^\pm$ as an alternative (but only one $Q=2$ term at once). 

The connection between these anomalous contributions and the underlying instanton-based physical picture can also be derived \cite{Rennecke:2020zgb}. It can be shown that $(\det\Phi)^Q$ is generated by multi-instantons with topological charge $Q$ and $(\det\Phi^\dagger)^Q$ by the corresponding anti--multi-instantons. Generally, higher-order contributions can also be included in $\mathcal{L}_{\rm qu}$. However, taking dimensional reduction at nonzero temperature as well as parity conservation into account, further $Q \geq 2$ terms are irrelevant. 

Note, the term $(\det\Phi^\dagger)^Q$ contains a residual $Z_{(Q N_f)}$ symmetry. In particular, the last term of Eq. \ref{Eq:Lqu} contains an $Z_2 \times Z_3 \cong Z_{6}$ symmetry, which is larger than the expected center symmetry $Z_{3}$ of the QCD vacuum in the chiral limit. This fact may point to a nonzero contribution of the first (and/or the second) term. Yet, symmetry arguments do not inform us about the relative strength of these contributions. 

\subsection{Parametrization of the eLSM}
In order to correctly implement the axial anomaly in vacuum, at least one of the anomalous couplings $\xi_1$, $\xi_1^{(1,1)}$, $\xi_2$, $\xi_2^{+}$, or $\xi_2^-$ has to be nonzero. Hence, if $\xi_1$ vanishes, the anomaly is encoded in higher-order correlations, as discussed in Ref.~\cite{Pisarski:2019upw}. In principle, it would be possible to incorporate temperature- and chemical potential-dependence of these couplings. On the other hand, this would require either additional assumptions on this dependence, or a more advanced, beyond mean-field treatment of the model. In the lack of these, we restrict ourselves to a self-consistent setup with constant parameters.

Following the argument of Ref. \cite{Pisarski:2024esv}, one can define an effective anomaly coupling including the contributions of the different terms in Eq.~\eqref{Eq:Lqu}. However, in the presence of the $Q=2$ terms, the leading field dependence cannot be properly factored, and hence different quantities show different relative contributions from these terms. From the expressions for $m_\pi^2$ and $m_{\eta_N}^2$ (see the appendix of Ref.~\cite{Giacosa:2024orp})---that are distinguished in the model only by the axial anomaly---the effective coupling
\begin{align} \label{Eq:chieff}
\xi_\text{eff}\equiv\xi_1 + \xi_1^1 \left(\phi_N^2+\phi_S^2\right)/2+\xi_2 \phi_N^2 \phi_S / \sqrt{2} \text{ }
\end{align}
can be deduced. This is expected to best reflect the relationship between the different couplings, since the $m_\pi^2$--$m_{\eta_N}^2$ mass splitting is controlled by the chiral anomaly. Due to the condensate dependence, $\xi_\eff$ becomes implicitly temperature and chemical potential-dependent. In chiral models, the contribution of the anomaly is always proportional to the---temperature and chemical potential dependent---chiral condensate, in line with general expectations \cite{Kwon:2012vb}.

To determine the model parameters, including each nonzero $\xi_X=\lbrace\xi_1,\xi_1^1,\xi_2,\xi_2^{\pm}\rbrace$, we fit physical quantities, such as meson masses and decay widths calculated in the model, to their experimental values (for a detailed description and the complete set of quantities see Fit$^{1,1,1,2}$ in Ref.~\cite{Kovacs:2016juc}). The parametrization with different combinations of the anomaly terms leads to two important conclusions. First, each combination of terms can provide a fit with a reasonable $\chi^2$ value, hence neither of them can be excluded solely based on vacuum phenomenology. Second, the anomaly-related parameters obey the constraint given by the expectation that only an effective anomaly coupling can be determined, which supports the observation in Ref.~\cite{Pisarski:2024esv}. This is shown in Fig.~\ref{Fig:parametrizations}, where the best 100 parameter sets are depicted for each setup. Note that in the individual sets the value of $\phi_N$ and $\phi_S$ vary, as well as the effective coupling $\xi_\eff$ might be slightly different. This leads to the fluctuation around the approximate joint surface.

We also note that here we considered not only the case where each coupling is restricted to be nonnegative, but also the general case without such restrictions. While from the point of view of the parametrization the individual sign of the couplings is irrelevant, it turns out that non-negligible negative couplings usually lead to an unstable solution at nonzero $T$ or even in the vacuum ($T=0$). Therefore, we use only parameter sets that have each $\xi_X>0$ (circles and triangles in Fig.~\ref{Fig:parametrizations}) in the subsequent calculations.

\subsection{Extending the anomaly terms within the eLSM: the polydeterminant}

A generalization of the determinant appears in effective Lagrangians modelling
the chiral anomaly in QCD when different types of mesons are considered \cite{Giacosa:2017pos,Giacosa:2023fdz,Giacosa:2024epf}. This \textit{polydeterminant} function, also known in the
mathematical literature as mixed discriminant, associates $N$ distinct
$N\times N$ complex matrices into a complex number and reduces to the usual
determinant when all matrices are taken as equal \cite{Giacosa:2025dvu}.

Given $N$ complex $N\times N$ matrices $A_{1},A_{2},...,A_{N},$ the
polydeterminant $Pdet=\varepsilon:%
%TCIMACRO{\U{2102} }%
%BeginExpansion
\mathbb{C}
%EndExpansion
^{N^{2}}\rightarrow%
%TCIMACRO{\U{2102} }%
%BeginExpansion
\mathbb{C}
%EndExpansion
$ reads:%
\begin{equation}
\varepsilon(A_{1},A_{2},...,A_{N})=\frac{1}{N!}\varepsilon^{i_{1}i_{2}%
...i_{N}}\varepsilon^{i_{1}^{\prime}i_{2}^{\prime}...i_{N}^{\prime}}%
A_{1}^{i_{1}i_{1}^{\prime}}A_{2}^{i_{2}i_{2}^{\prime}}...A_{N}^{i_{N}%
i_{N}^{\prime}}\text{ ,}%
\end{equation}
where $i_{k}=1,2,...,N$ and $i_{k}^{\prime}=1,2,...,N$. It is easy to prove
that this function is linear and that various properties exist, e.g.:
$\varepsilon(A,A,...,A)=\det\left(  A\right)  $, $\varepsilon
(A,1,...,1)=Tr\left(  A\right)  ,$ $\varepsilon(BA_{1},BA_{2},...BA_{N}%
)=\det(B)\varepsilon(A_{1},A_{2},...A_{N})$, etc. We refer to Ref. \cite{Giacosa:2025dvu}
for a detailed list and related proofs.

In QCD with 3 flavors, one sets $N=N_{f}=3.$ For instance, if besides the
matrix $\Phi$ encountered in the previous section we also consider the
analogous matrix of excited (pseudo)scalar meson $\Phi_{E}$ (see Ref.
\cite{Parganlija:2016yxq} for their phenomenology) the anomalous terms analogous to $\det\Phi$
are:%
\begin{equation}
\varepsilon(\Phi,\Phi,\Phi_{E})\text{ },\varepsilon(\Phi,\Phi_{E},\Phi
_{E})\text{ },\varepsilon(\Phi_{E},\Phi_{E},\Phi_{E})=\det\Phi_{E}%
\end{equation}
Under chiral transformation $\Phi\rightarrow U_{L}\Phi U_{R}^{\dagger}%
,\Phi_{E}\rightarrow U_{L}\Phi_{E}U_{R}^{\dagger},$ thus these terms are
chirally symmetry but $U(1)_{A}$ violating. Such terms were not yet
investigated in a quantitative way: this is left for the future. 

One may also consider chiral multiplets with nonzero spin. Again, one should
consider heterochiral multiplets, that is chiral multiplets transforming as
$\Phi\rightarrow U_{L}\Phi U_{R}^{\dagger}$, see the classification in Refs.
\cite{Giacosa:2017pos,Giacosa:2024epf}. Two examples are given by the heterochiral (pseudo)vector mesons
$\Phi_{\mu}$ (involving the $1^{-+}$ pseudovector, such as the state
$b_{1}(1235)$ and their flavor and chiral partner, the $1^{--}$ radially
excited vector mesons, such as $\rho(1700)$ \cite{Piotrowska:2017rgt}), and by the heterochiral
pseudotensor mesons $\Phi_{\mu\nu}$ (involving  $2^{-+}$ states such as
$\pi_{2}(1880)$ \cite{Koenigstein:2016tjw}). Due to the chiral transformations $\Phi_{\mu
}\rightarrow U_{L}\Phi_{\mu}U_{R}^{\dagger}$ and $\Phi_{\mu\nu}\rightarrow
U_{L}\Phi_{\mu\nu}U_{R}^{\dagger},$ Lorentz-invariant chirally symmetric but anomalous interactions take the form
\begin{equation}
\varepsilon(\Phi,\Phi_{\mu},\Phi^{\mu})\text{ , }\varepsilon(\Phi,\Phi_{\mu
\nu},\Phi^{\mu\nu}),...
\end{equation}
Mixing patterns in the isoscalar sector (formally similar to the famous mixing
leading to $\eta(547)$ and $\eta^{\prime}(958)$) and specific decay channels
follow from such interactions. The strength of these interaction terms can be estimated using a dilute instanton gas, in a way similar to the pseudoscalar case, see Ref. \cite{Giacosa:2023fdz}. 

Summarizing, the polydeterminant allows to introduce novel chirally anomalous interaction terms
among different types of mesons, thus allowing the extension of chiral models in general and the eLSM in particular along unexplored directions.

%%%%%%%%%%%%%%%%%%%%%%%%%%%%%%%%%%%%%%%%%%
%%%%%%%%%%%%%%%%%%%%%%%%%%%%%%%%%%%%%%%%%%
\section{The Columbia plot} \label{sec:columbia}
%%%%%%%%%%%%%%%%%%%%%%%%%%%%%%%%%%%%%%%%%%
%%%%%%%%%%%%%%%%%%%%%%%%%%%%%%%%%%%%%%%%%%

%%
%%%%%%%%%%%%%%%%%%%%
\begin{figure*}[t]
    \centering
    \includegraphics[width=.32\textwidth]{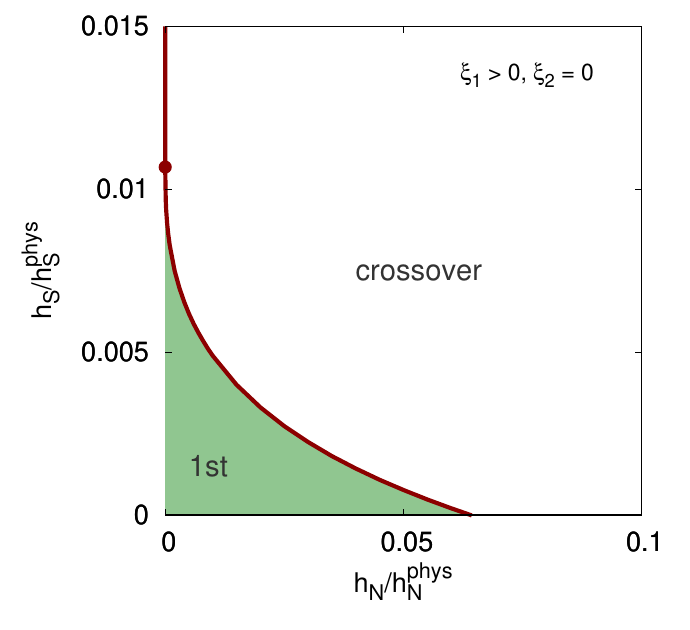}
    \includegraphics[width=.32\textwidth]{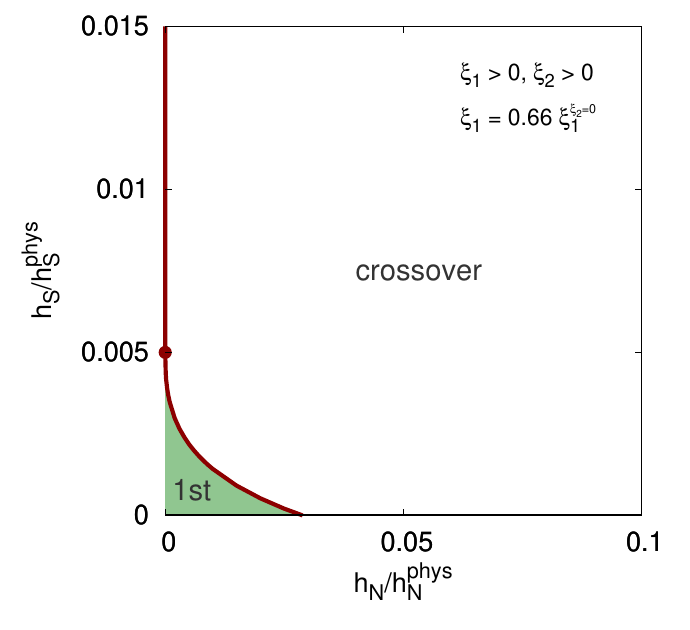}
    \includegraphics[width=.32\textwidth]{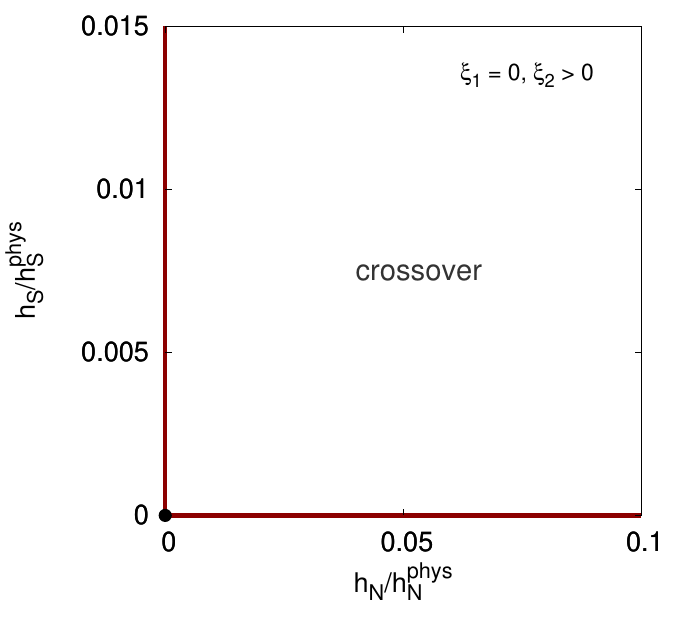}
    \caption{The Columbia plot in the plane of the normalized explicit symmetry breaking parameters with parameters including $\xi_1$ (left), $\xi_2$ (right), and both (center).}
    \label{Fig:Columbia_plot_xi1xi2}
\end{figure*}
%%%%%%%%%%%%%%%%%%%%
%%

After carrying out the vacuum study in Sec. 2.2, viable sets of parameters regarding vacuum phenomenology are available, and the resulting phase transition can be explored. We solve the field equations at nonzero $T$ to obtain the chiral condensates $\phi_{N,S}$ and hence determine the order of the phase transition. Repeating this procedure at different values of the explicit symmetry-breaking parameters, $h_N$ and $h_S$  we can draw the Columbia plot within our model. 

The square mass of the pion $m_{\pi}^2 = Z_{\pi}h_N/\phi_N$ suggests that $h_N$ is proportional to the bare quark mass $m_n =(m_u+m_d)/2$. Similarly, $h_S \propto m_S$.
Yet, assuming a strict simple proportionality is valid only at leading order.  In fact, the parameters of the low energy model can be extracted by integrating out the fluctuation between $\Lambda_{\rm UV} \gg 1\,{\rm GeV}$, a fixed high-energy scale in the perturbative regime where the masses are fixed, and $\Lambda_{\rm eLSM}\lesssim 1-2\,{\rm GeV}$, which is the upper scale of validity of the model \cite{Rennecke:2015lur, Springer:2016cji, Fu:2019hdw}. The different quark masses at $\Lambda_{\rm UV}$ lead to different values for all model parameters, not only for $h_N$ and $h_S$. At the physical point, these parameters can be determined by comparing to experimentally measured hadron properties, which we discussed in the previous section. However, as we move away from the critical point by changing $m_{u,d}$ and $m_s$, the model parameters modify in a nontrivial way.

A large-$N_c$ analysis \cite{tHooft:1973alw,Witten:1979kh} may help to clarify the issue above and to present a concrete example. The scaling $h_N \sim N_c^{1/2}$ is required, which can be obtained by modifying the ESB as \cite{Giacosa:2024scx}:
\begin{equation}
    \mathcal{L}_{\rm ESB}=\lambda G^2\tr(M\Phi)
    \text{ ,}
\end{equation}
where $G$ is the glueball/dilaton field and $M = diag(m_u,m_d,m_s)$ contains the bare quark masses. The parameter $\lambda \sim N_c^{-3/2}$ because it describes a glueball-glueball-$q\bar{q}$ interaction. Then, in the isospin-symmetric limit, the parameter $h_N =\lambda m_nG_0^2$ correctly scales as $N_c^{1/2}$, but the minimum of the dilaton field $G_0$ is also dependent on $m_n$ \cite{Janowski:2014ppa}. We may conclude that $h_N =m_n f(m_n,m_s)= m_n f(0,0)+...$. Thus, only at leading order, $h_N$ is simply proportional to $m_n$.

A full treatment of this problem---similar to that in chiral perturbation theory \cite{Herpay:2005yr, Resch:2017vjs} to circumvent the lack of fundamental QCD input---would be extremely complicated in the eLSM, and is beyond the scope of the present work\footnote{For completeness and safety, we restrict our parameters to provide a description qualitatively compatible with chiral perturbation theory results.}. Therefore, we concentrate on the scenario where only $h_N$ and $h_S$ are varied without mapping them exactly to $m_{u,d}$ and $m_s$. At leading order and for the qualitative purposes of this work, this is sufficient to derive the Columbia plot.

In Fig.~\ref{Fig:Columbia_plot_xi1xi2} we show the resulting Columbia plots for three different, representative parameter sets. Note that we only show results with varying $\xi_1$ and $\xi_2$, while keeping $\xi_1^{(1,1)} = 0$. Contrary to Ref.~\cite{Giacosa:2024orp}, we use $h_S/h_S^\phys$ and $h_N/h_N^\phys$ instead of $m_K$ and $m_\pi$, which causes a deformation of the phase structure for $h_S>0$ that is presented here.

The left panel shows the Columbia plot where the anomaly is implemented only with the conventional 't~Hooft determinant, i.e.,~$\xi_1 = \bar\xi_1 > 0$ and $\xi_2 = 0$. This is in qualitative agreement with other mean-field results where the anomaly is implemented in the same way, e.g.,~\cite{Resch:2017vjs}. In the $h_N=0$ chiral limit, and physical strange quark mass, $h_S = h_S^{\rm phys}$, we find a second-order transition. This extends towards smaller $h_S$ until it ends in a tricritical point. Below this point, an extended region where the chiral transition is of first order is found. Starting from the tricritical point, a second-order transition continues with decreasing $h_S$ and increasing $h_N$ until it ends at $h_S = 0$ and $h_N > 0$. For $h_N = 0$ and even larger $h_S$, only a crossover is found. This is consistent with the expectation for the one-flavor limit, which is reached for $h_S = 0$, $h_N \rightarrow \infty$. In this case, $\xi_1$ explicitly breaks chiral symmetry, leading to a nonzero chiral condensate for all temperatures. 

When $\xi_2 > 0$ is considered, the tricritical point on the left edge moves to lower $h_S$ values. Consequently, the first-order region near the $N_f=3$ chiral limit shrinks as $\xi_2$ is increasing and---due to $\xi_\eff$ being approximately constant among the parametrisations---$\xi_1$ is decreasing. This is shown in the center panel, where $\xi_1=0.66\,\bar\xi_1$ is used. As $\xi_1$ further decreases, the first-order region shrinks and disappears at $\xi_1=0$, as can be seen on the right panel. In this scenario, the transition becomes second-order also in the $h_N>0$, $h_S=0$ chiral limit, and the Columbia plot at $h_N\geq0$ and $h_S\geq0$ resembles that in the complete absence of the $U(1)_A$ anomaly. Hence, only from the this picture, the $\xi_\eff\propto \xi_2$ and the $\xi_\eff=0$ cases cannot be distinguished. Interestingly, this second-order line at $\xi_1=0,~h_S=0$ is continuously connected to the second-order line at $\xi_1>0$,~$h_S>0$, just as the specially symmetric point at $h_S=h_N=0$ is connected to the tricritical point. 
Since the presence of the first-order region is favored only in the case of $\xi_1>0$ (each $\xi_2$, $\xi_2^\pm$, and $\xi_1^1$ suppresses this region), it can be concluded that the relative strength of the 't~Hooft coupling $\xi_1/\xi_\eff$ governs the size of this region. In light of recent lattice results, this suggests that the $\xi_1$ term might not be the dominant contribution for the axial anomaly. Yet, if we insist that the center symmetry should not be larger than $Z_3$, the parameter $\xi_1$ should not vanish: an eventually small but nonzero first-order region is still present in the left-down part of the Columbia plot. 

Finally, we note that at $h_S<0$, which was already partially presented in the $m_K\ll m_\pi$ region in Ref.~\cite{Giacosa:2024orp}, the phase structure is more complicated, possibly including CP-violating phases. This interesting region will be discussed in a future publication.

%%%%%%%%%%%%%%%%%%%%%%%%%%%%%%%%%%%%%%%%%%
%%%%%%%%%%%%%%%%%%%%%%%%%%%%%%%%%%%%%%%%%%
\section{Conclusions}
%%%%%%%%%%%%%%%%%%%%%%%%%%%%%%%%%%%%%%%%%%
%%%%%%%%%%%%%%%%%%%%%%%%%%%%%%%%%%%%%%%%%%
In this work, we reported on some features related to the chiral anomaly, both in vacuum and in the medium. For the ground-state (pseudo)scalar sector, we included terms proportional to the standard 't Hooft determinant term(s) generated by instantons with topological charge one, and to the square of the determinant, which emerges from instantons with topological charge 2. Both terms are consistent with vacuum phenomenology (Fig. 1). This additional term turns out to be important when studying the Columbia plot at nonzero temperature \cite{Giacosa:2024orp}. In particular, the first-order region in the low-left corner of the diagram shrinks when this term dominates (Fig. 2).

When other chiral multiplets are considered, an analogous anomalous 't Hooft term can be introduced by using an extension of the determinant, the so-called polydeterminant \cite{Giacosa:2017pos,Giacosa:2023fdz,Giacosa:2025dvu}. A new class of interaction is possible, whose detailed study represents an outlook of this work.

\bigskip
\textbf{Acknowledgement}
%We acknowledge discussions with Christian Fischer, Owe Philipsen, Bernd-Jochen Schaefer, Lorenz von Smekal, Gy\"orgy Wolf,...
%F.R.\ is supported by the Deutsche Forschungsgemeinschaft (DFG, German Research Foundation) through the Collaborative
%Research Center TransRegio CRC-TR 211 “Strong- interaction matter under extreme conditions” – project number 315477589 – TRR 211.
The authors thank M. Stephanov for useful discussions. 
P. K. and G. K. acknowledge support by the Hungarian National Research, Development and Innovation Fund under Project No. K 138277. The work of G. K. is partially supported by the Polish National Science Centre (NCN) under OPUS Grant No. 2022/45/B/ST2/01527.

\bibliography{columbia}

%\appendix*
%\input{sections/appendix1.tex}

\end{document}